\documentclass[twocolumn,amsmath,amssymb]{revtex4}
\usepackage{graphicx}

\begin{document}

\title{Measurement of Electron Backscattering\\ in the Energy
Range of Neutron $\beta$-Decay}

\author{J. W. Martin$^1$, J. Yuan$^1$, S. A. Hoedl$^2$, B. W. Filippone$^1$,
D. Fong$^1$,\\
T. M. Ito$^1$,  E. Lin$^1$, B. Tipton$^1$, A. R. Young$^3$}
\affiliation{$^1$W. K. Kellogg Radiation Laboratory,
California Institute of Technology, Pasadena, CA 91125\\
$^2$Princeton University, Princeton, NJ 08544\\
$^3$North Carolina State University, Raleigh, NC 27695\\
}

\begin{abstract}
We report on the first detailed measurements of electron
backscattering from low $Z$ targets at energies up to 124 keV. Both
energy and angular distributions of the backscattered electrons are
measured and compared with electron transport simulations based on the
Geant4 and Penelope Monte Carlo simulation codes. Comparisons are also
made with previous, less extensive, measurements and with measurements
at lower energies.
\end{abstract}

\maketitle

\section{Introduction}

Backscattering of electrons from the surfaces of bulk materials has
been studied at low energies largely in relation to materials science
applications (e.g. Auger electron spectroscopy and scanning and
transmission electron microscopy). For this reason, most detailed
studies have been conducted for incident electron energies less than
40 keV~\cite{bib:gerard,bib:massoumi1,bib:massoumi2}. At higher
energies ($E >>$ 1 MeV) sophisticated Monte Carlo calculations
exist~\cite{bib:geant4nim} that are constrained to reproduce data
obtained for nuclear or particle physics applications. However in the
intermediate regime ($0.04~{\rm MeV} < E < 1~{\rm MeV}$) there is
little data to constrain the simulations. This intermediate energy
regime can be important for various beta spectroscopy applications, in
particular neutron beta-decay ($E_{max} = 0.782~{\rm MeV}$).

Most of the data that exists in this energy region is based on
measurements using electron beams and detecting the backscattering
from bulk targets using electrical currents in Faraday cups (for a
recent review, see Ref.~\cite{bib:Tabata2}).  In particular, for
carbon and aluminum targets, extensive measurements using this
technique exist.  We have made measurements for beryllium and silicon
targets using this technique, and have investigated the effects of
secondary electron emission on these types of measurements.  In
addition, we have conducted measurements using ion-implanted silicon
detectors to detect the energy and angle of backscattered electrons,
extending the work of
Refs.~\cite{bib:gerard,bib:massoumi1,bib:massoumi2} to higher
energies. In this way, we can evaluate the reliability of existing
models to reproduce the dependence of the backscattering process on
energy and emission angle of the backscattered electrons.

The results of these measurements can have a variety of nuclear and
particle physics applications.  For example, an upcoming measurement
of the electron correlation in polarized neutron decay~\cite{bib:ucna}
will require an understanding of electron backscattering at the
10-20\% level.  In this experiment plastic scintillator and silicon
multi-strip detectors will be used, thus requiring detailed
information on backscattering for low $Z$ targets. As a first step in
addressing this problem, we have carried out detailed measurements of
the energy and angular distributions of backscattered electrons from
light materials (Be and Si) for incident electron energies between
43.5 and 124 keV.

\section{Experiment Overview}

The experiment consisted of an electron gun and a scattering chamber
containing a movable target.  Two modes of running were used where
backscattering data could be acquired.  In one mode the silicon
detector was used to detect the energy and angle of backscattered
electrons.  In a second higher-current mode the electrical current due
to the backscattered electrons incident on the chamber walls was
detected.  These two modes will be referred to as silicon detector
mode and current integration mode, respectively.  Each component of
the experiment will be described in greater detail.

\subsection{The Electron Gun}

\begin{figure*}
\includegraphics[width=\textwidth]{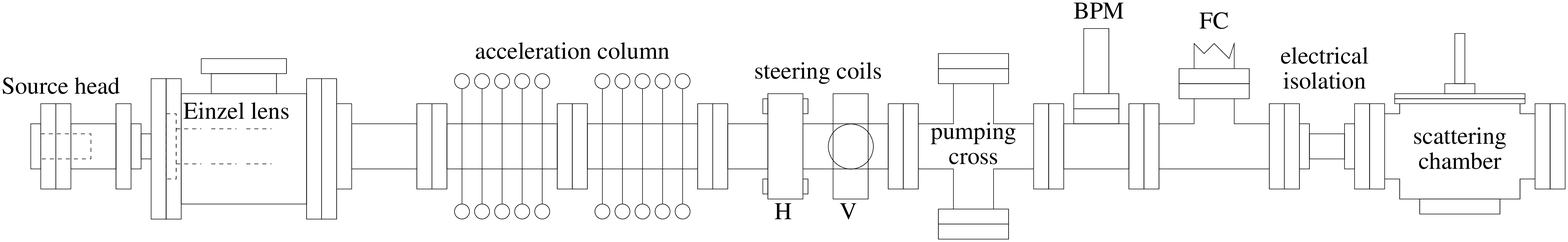}
\caption{Schematic of the Kellogg Electron Gun.  BPM indicates the
location of the beam position monitor.  FC indicates the location of
the Faraday cup.  H and V indicate the horizontal and vertical
steering coils, respectively.\label{fig:gun}}
\end{figure*}

The Kellogg Electron Gun was constructed in order to perform these
measurements.  A schematic of the gun is shown in
Fig.~\ref{fig:gun}.
Electrons were extracted from a hot filament by a potential of
typically 6 kV, in the source head.  The electrons were focused by an
Einzel lens before passing through an additional acceleration
potential of up to 130 kV minus the extractor voltage.

The electron beam current could be varied from a few electrons per
second to several $\mu$A, by adjusting the current passed through the
filament.  The resultant beam current remained stable after allowing
the filament to warm up for typically half an hour.

After the acceleration column, two magnetic coils allowed steering of
the beam in the horizontal and vertical directions.

A Faraday cup could be inserted into the beam to monitor the total
current of the beam.  A beam position monitor, consisting of a
rotating wire which sweeps through the beam twice along two orthogonal
axes, could be used to locate the beam within the beam pipe.

The energy of the electron gun was stable and reproducible.  The
energy was absolutely determined to the 1\% level using a novel
Helmholtz coil spectrometer.  This spectrometer is iron-free and has a
momentum resolution of 0.3\%~\cite{bib:junhua}.  The energy of the
electron gun was found to be monochromatic to better than the
resolution of the spectrometer.

\subsection{Chamber, Targets and Detector}

\begin{figure}
\includegraphics[width=0.5\textwidth,angle=270]{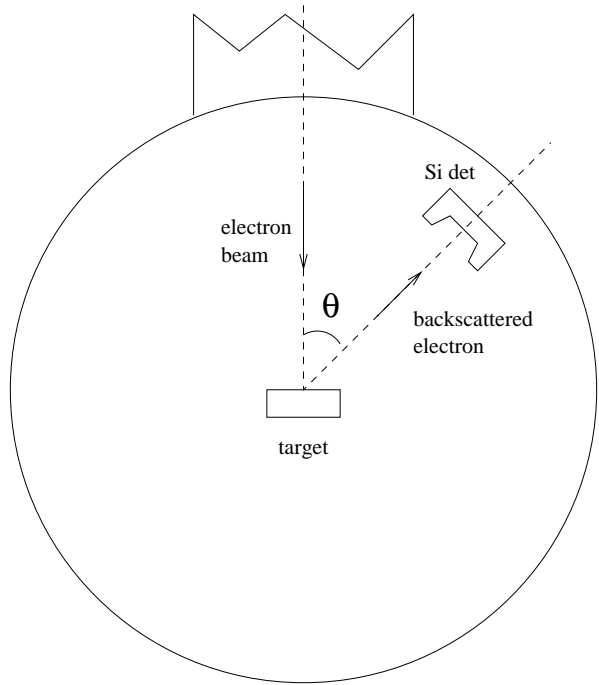}
\caption{Schematic of the Chamber and Detector\label{fig:chamber}}
\end{figure}

A schematic of the chamber and detector arrangement used to perform
the backscattering measurements is shown in Fig.~\ref{fig:chamber}.
The chamber contained a target and a silicon detector on separately
rotatable feedthroughs.  The chamber, target, and detector were
each separately isolated from ground.  The electric currents detected by
the chamber and the target were integrated using two Ortec model 439
current digitizers.

The targets~\cite{bib:goodfellow} were multiple stopping lengths of
beryllium and silicon.  Two different beryllium targets of nominal
thicknesses 1.0~mm and 2.0~mm were used.  Two different silicon
targets of thickness 1.0~mm were used.  The targets were 25~mm by
25~mm square, and were held at one corner in a slotted rod by a set
screw.  The beryllium targets were nominally 99.0\% pure, and had a
metallic appearance and the polycrystalline silicon target was
nominally 99.999\% pure.

The energy and angular distribution of the backscattered electrons was
measured using an Ortec ion-implanted silicon detector (Ortec model
BU-13-25-300, 25 mm$^2$ nominal active area and 300 $\mu$m nominal
depletion depth at 100 V bias).  The detector was mounted on a
rotatable arm which allowed it to be placed at an arbitrary angle with
respect to the target.  The target to detector distance was typically
8.5 cm.  The detector linearity and resolution was calibrated by
placing the detector in the electron beam at very low beam current
($\sim$ 100 Hz).  The detector linearity was confirmed to the
sub-kilovolt level for these experiments.  The energy resolution of
the silicon detector was found to be typically $\sigma=2.5$ to
3.5~keV, and the resolution was independent of energy to less than 0.1
keV.

For fixed filament current, the current detected on the target was
maximized in order to tune the beam. The beam tunes for each energy
were stored on the computer that controlled the power supplies of the
accelerator. In addition, tunes were checked using a scintillator
target coated with graphite which produced a visible spot ($\sim$ 3~mm
$\times$ 3~mm) indicative of the beam spot size when struck with
sufficient current (typically 1 nA).

\section{Backscattering Measurements}

Backscattering measurements were performed for normal incidence upon
the target.  Measurements were performed for incident electron
energies of 43.5, 63.9, 83.8, 104, and 124 keV.  For each energy, both
silicon detector mode and current integration mode measurements were
taken.

\subsection{Silicon Detector Mode}

In the silicon detector mode, silicon detector spectra were acquired
for backscattered electron angles of 20 to 80 degrees in steps of 10
degrees.  The backscattered angle $\theta$ was defined with respect to
the normal of the target, as shown in Fig.~\ref{fig:chamber}.

When the beam was off, the detector rate was typically 100 Hz, due to
low energy noise.  The detector rate with beam on was typically 5 kHz,
and was always kept below 20 kHz, to limit ADC pile-up to below 3\% on
average across the spectrum.  The beam current for each target and
detector angle was therefore different, due to the different levels of
backscattering.  For Si, the current was typically 40 pA, while for
Be, it was 300 pA.

\begin{figure}
\includegraphics[width=0.5\textwidth]{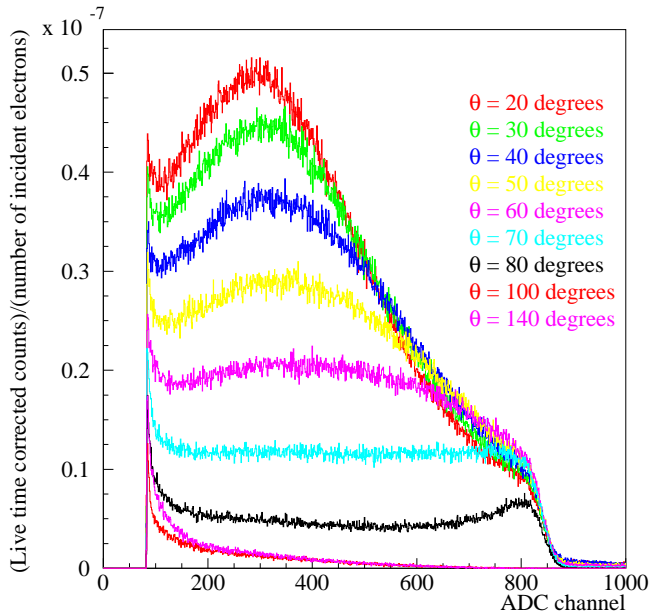}
\caption{(Color online) Silicon detector raw spectra for normal
incidence backscattering from beryllium target at $E_{\rm
beam}=124$~keV.\label{fig:raw}}
\end{figure}

Fig.~\ref{fig:raw} shows normalized spectra taken using the silicon
detector for a variety of detector angles $\theta$ for 124~keV
electron beam energy and the beryllium target.
The data are plotted as a function of the digitized pulse height (ADC
channel), which is proportional to the energy $E$ deposited in the
silicon detector.  The data are not corrected for the response of the
silicon counter, which had a resolution of about 2.5 keV, and itself
suffered from backscattering. Below about 20 keV, contributions from
noise in the detector and electronics contributed to the count rate at
the few percent level, and so no data is shown beneath that threshold.

A detector rate above electronic noise was also detected for angles
$\theta>90^\circ$, where the detector is shielded from any direct
backscattering.  It can be seen that the rates at $100^\circ$ and
$140^\circ$ are significant, especially compared to e.g. the
$80^\circ$ measurements.  The flux found at $\theta>90^\circ$ was
determined to be due to rescattering from the walls of the chamber,
and was confirmed in Monte Carlo studies.  In these studies,
backscattering from the steel walls and lid of the chamber were
included.  The chamber floor, dominantly aluminum, had relatively
little effect on the chamber background and was not included. The
Monte Carlo calculations showed that electrons at $\theta>90^\circ$
could be produced, and that the result was numerically about 50\%
larger than the observed chamber background.  The disagreement is
attributed to the use of a simplified geometry in the Monte Carlo, and
uncalibrated treatment of backscattering from these sources.

Motivated by the simulation, experimental studies were carried out
lining the chamber walls with stopping thicknesses of aluminized mylar (which
has a much lower backscattering fractions).
This was found to reduce the chamber background by roughly 50\%, a
figure in agreement with expectations based on the Monte Carlo.

Based on the Monte Carlo simulations, the $100^\circ$ data is expected
to represent most closely the chamber background for the $80^\circ$
data, and to represent an upper bound on the chamber background for
the other angles.  So the $100^\circ$ data was subtracted from the
$\theta=20^\circ-80^\circ$ data to arrive at the background-subtracted
spectrum.  The systematic error in using this subtraction scheme is
3-5\% for 124 keV beam energy, peaked at low energy deposition, and is
smaller for lower beam energy (as more of the chamber background is
moved into the unmeasured region dominated by detector noise).

From Monte Carlo studies and analytical estimations, it was also
determined that contributions from X-rays were negligible in both flux
(compared to electron backscattering) and in detection efficiency (for
the 300 $\mu$m thick detector used).  Therefore no corrections were
made for X-rays.

\subsection{Systematic uncertainties for silicon detector 
measurements\label{sec:syst}}

Table~\ref{tab:sisyst} summarizes the systematic uncertainties for the
silicon detector measurements.
\begin{table}[ht]
\begin{tabular}{|l|c|}\hline
Effect & Uncertainty \\\hline
reproducibility & 7\% \\
active area & 4\% \\
beam spot size & 5\%$\times\sin\theta$ \\
deadtime & 3\%\\
alignment & 2\%\\
current detection & 3\% \\
\hline
Total & 12\% average\\
\hline
\end{tabular}
\caption{Summary of systematic uncertainties associated with silicon
detector measurements.  The total systematic uncertainty ranges from
11\% to 15\%, depending on angle, averaging 12\%.\label{tab:sisyst}}
\end{table}
Each effect in the table will now be described in more detail.

Measurements of the backscattered electron yield in the silicon
detector were found to be reproducible at the 7\% level.  The detector
active area was measured using an alpha source and various
collimators, and found to be consistent with the geometry of the
detector to about 4\%.  The beam spot size and detector size were also
measured by scanning the beam across the detector.  The detector size
was again found to be in agreement with the directly measured value.
At the same time, it was observed that no significantly different
response for electrons could be seen for the detector as one varied
the position of the beam on the detector surface.  A simple silicon
response function including only normal-incidence backscattering was
therefore used to convolute Monte Carlo predictions for comparison
with the data (see section~\ref{sec:MC}).  The beam spot radius was
found to be about 1.7~mm, and the beam was always well-centered on the
target to this level, limiting a solid angle correction from finite
spot size to less than $5\%\times\sin\theta$.  Deadtime corrections
varied between $< 10\%$ and $50\%$.  However, ADC pile-up was kept
below 3\% and the deadtime corrections could be performed reliably at
that level, using information from fast scalers counting the triggers.

The relative alignment of the beam, target, and detector in the plane
of rotation of the detector was measured to the level of 0.5
degrees.  This was confirmed by taking measurements of electron
backscattering at both positive and negative $\theta$.  The residual
systematic error due to alignment problems was estimated to be about
2\%.

The silicon measurements were normalized to the total current detected
by the chamber and target.  The detection of current was calibrated
using a precision picoampere source, and comparison among several
well-calibrated picoammeters.  The calibration agreed each time at the
level 0.3 pA, giving a worst case contribution to the normalization
uncertainty of 3\%.  Integration of low-energy secondary electrons was
not a factor for this measurement of current, since the total charge
on the chamber and target was collected.

The total normalization systematic uncertainty after background
subtraction was therefore 12\%, dominated by reproducibility and
detector solid angle uncertainties.

\subsection{Current Integration Mode}

For each energy, current integration mode measurements were also
performed.  By separately sensing the currents on the target ($I_{\rm
target}$) and the chamber ($I_{\rm chamber}$), the total
normal-incidence backscattered fraction $\eta$ was determined via:
\begin{equation}
\eta=\frac{I_{\rm target}}{I_{\rm target}+I_{\rm chamber}}.
\end{equation}
Low energy secondary electrons may be emitted from the surfaces of
materials after higher energy electrons strike the surface.  In
previous experiments, secondary electrons were typically defined to
have $E<50$~eV.  If the fraction of secondary electrons produced per
high-energy interaction is large, the sensed currents will give an
erroneous measure of the high-energy backscattered fraction.

To suppress and quantify the effects of secondary electrons, a
cylindrical ``cage'' of wires (referred to as the grid) was inserted
into the setup to provide a potential wall between the chamber and the
target.  This grid was made of 50~$\mu$m tungsten wire, wrapped on a
cylindrical copper frame of radius 6~cm and height 8~cm.  The grid had
22 vertical wires evenly spaced on the sides.  The wires met at the
bottom, but had an opening at the top so that the target rod (which
holds the target) could be inserted.  The grid, target, and chamber
could each be biased at different voltages up to a difference of 200~V
with leakage currents between elements kept to below 1~pA.

\subsection{Systematic Uncertainties for Current Integration Measurements}

Table~\ref{tab:currsyst} summarizes the systematic uncertainties
encountered for the current integration measurements.
\begin{table}[ht]
\begin{tabular}{|l|c|}\hline
Effect & Uncertainty \\\hline
target rod correction & 7\% \\
grid secondaries & 1\% \\
reproducibility & 5\% \\
current dependence & 3\%\\
\hline
Total & 9\% \\
\hline
\end{tabular}
\caption{Summary of systematic uncertainties associated with current
integration measurements.\label{tab:currsyst}}
\end{table}
Each of these uncertainties will now be described in more detail.

It was found that biasing the grid at roughly -50~V relative to the
chamber and target caused secondaries created on the chamber and
target to be recollected by the chamber and target.  When the target
was biased to +50~V, and the chamber held at ground, the effect of
changing the grid voltage from zero to -50~V resulted in changes of
10-30\%, depending on the target material and the beam energy.  This
agreed well with estimates of the effect based on the Penelope Monte
Carlo code, and based on measurements of secondary electron emission
summarized in Refs.~\cite{bib:joy,bib:bronshtein,bib:bronshtein2}.

Due to a small piece of the conducting target rod (held at the same
potential as the chamber) penetrating into the top of the grid, there
was a residual correction still to be made for secondaries.  In the
data, this showed up as a residual dependence of $\eta$ on the
relative target/chamber voltage, even when the grid was set to very
large voltages.  It was found that a correction could be made using
the solid angle subtended by that piece of the target rod, and the
value of $\eta$ determined when the grid was not used.  This
correction amounted to a 7\% contribution to the systematic
uncertainty in the determination of $\eta$.

Secondaries due to high-energy electrons striking the grid could be
accounted for at the 1\% level.  Measurements of $\eta$ under widely
varying beam conditions showed that it was reproducible at the 5\%
level.

The current dependence of $\eta$ was also studied from pA to $\mu$A in
incident beam current, and found to vary at the 3\% level.  This
indicated that electrostatic charging of various non-conducting
components in the setup contributed at a small level.

The total fractional systematic uncertainty in $\eta$ determined from these
current-mode measurements was therefore 9\%.

\section{Results}

\subsection{Silicon Detector Mode}

\begin{figure}
\includegraphics[width=0.5\textwidth]{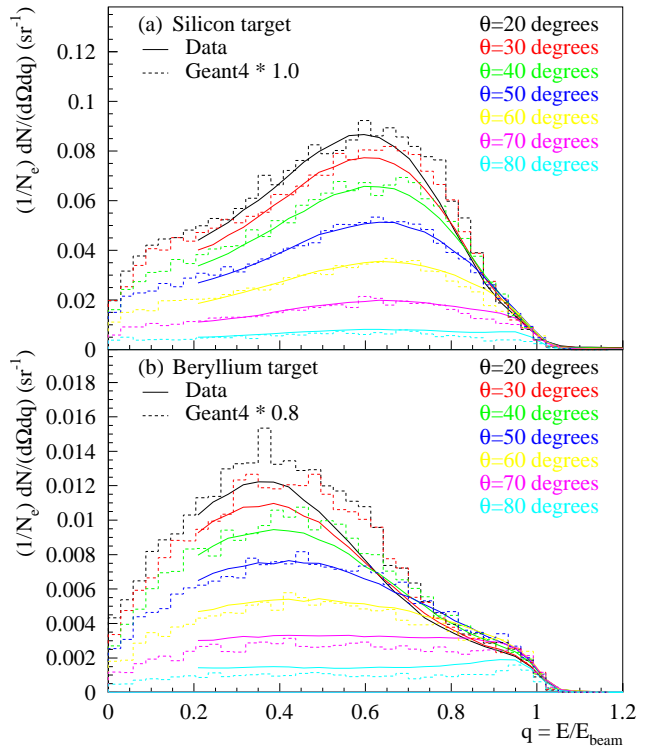}
\caption{(Color online) Normal incidence backscattering from
a. silicon, and b. beryllium targets at $E_{\rm beam}=124$~keV.
Curves represent rebinned data taken with silicon detector.  Histogram
is Monte Carlo simulation based on Geant4.  Systematic uncertainty in
the normalization of the data is estimated to be 12\% on average,
ranging from 11\% at small angles to 15\% at large angles.  For
beryllium, a scale factor of 0.8 is applied to the Monte Carlo
simulation, while for silicon, no Monte Carlo scale factor is
applied.\label{fig:120}}
\end{figure}

The normalized, background-subtracted spectra accumulated for various
detector angles for 124~keV electrons normally incident on silicon and
beryllium bulk targets are shown in Fig.~\ref{fig:120}.
The Monte Carlo curves will be described in Section~\ref{sec:MC}.  The
data are plotted as a function of the dimensionless energy $q=E/E_{\rm
beam}$, where $E$ is the energy detected by the silicon detector and
$E_{\rm beam}$ is the energy of the incident electrons, in this case
124~keV.  On the vertical axis, 
$\frac{1}{N_e}\frac{dN}{dqd\Omega}$, the number of counts per incident
electron, per unit $q$, per unit solid angle is plotted.  
In the absence of the
effects of detector response (resolution and backscattering), this
would be the normal-incidence backscattered fraction per unit $q$, per
unit solid angle.

In Fig.~\ref{fig:120}a., for silicon, for small backscattered angles
($\theta=20^\circ-30^\circ$), a peak is found near $q=0.65$, and a
shoulder found near $q=0.95$.  As the backscattered angle increases,
the peak at $q=0.65$ tends to disappear and shift slightly to higher
$q$, while the shoulder at $q=0.95$ tends to become more pronounced as
events at lower $q$ disappear.

The same trends can be seen in the spectra for a beryllium target,
shown in Fig.~\ref{fig:120}b.  However, in this case, the low-energy
peak appears closer to $q=0.35$.  This can be explained by the
beryllium having smaller Rutherford scattering cross section.
Electrons therefore penetrate more deeply into the material before
scattering.

The silicon results compare well qualitatively with the results of
Refs.~\cite{bib:gerard,bib:massoumi1,bib:massoumi2}, which were
acquired at lower energies on silicon and aluminum targets.  It is
difficult to compare directly with these measurements.  For
Ref.~\cite{bib:massoumi1,bib:massoumi2}, no data is published above 35
keV and more data are displayed for oblique incidence.
Ref.~\cite{bib:gerard}, while providing data at 40 keV, gives no
absolute normalization.

\begin{figure}
\includegraphics[width=0.5\textwidth]{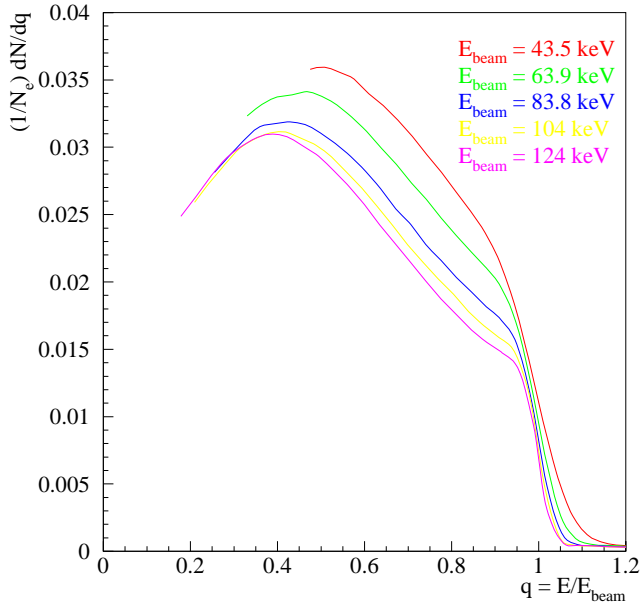}
\caption{(Color online) Normal incidence backscattering from beryllium
target at each beam energy, integrated over angles.  Systematic
normalization uncertainty is 13\% in each case.\label{fig:be50deg}}
\end{figure}

The dependence on beam energy was also investigated.
Fig.~\ref{fig:be50deg} shows the dependence of the backscattered
fraction on both the beam energy and the energy of the backscattered
electron, integrated over all possible backscattered angles.
The integration over angle was performed using a finite sum with
$10^\circ$ bins centered on each measured angle, and the appropriate
solid angle weighting.  A small correction due to the unmeasured
regions at small and large angles was included.  The systematic
uncertainty in approximating the integral by the sum was typically
4\%, from comparison with analytical forms.  Angular bin centering
corrections were found to be negligible, due to the smoothness of the
angular behavior.  The overall systematic uncertainty was consequently
increased to 13\%.

As seen in Fig.~\ref{fig:be50deg}, when plotted in terms of the
dimensionless variable $q$, the curves nearly overlap.  The same
qualitative behavior can be seen as a function of beam energy for
individual backscattered angles. The data are therefore observed to
follow a near-scaling behavior.  The overall normalization follows the
energy dependence of the total normal-incidence backscattered
fraction.

\begin{figure}
\includegraphics[width=0.5\textwidth]{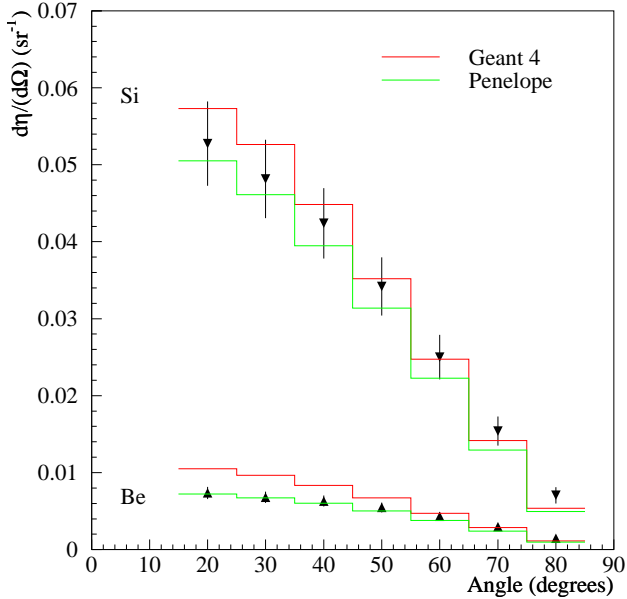}
\caption{(Color online) d$\eta$/d$\Omega$ for beryllium (triangles)
and silicon (inverted triangles) targets at $E_{\rm beam}=124$~ keV.
Black points with error bars indicate data with total normalization
systematic uncertainties shown.  Red histogram indicates the results
of the Geant4-based Monte Carlo simulation.  Green histogram indicates
the results of the Penelope-based Monte Carlo simulation.  No Monte
Carlo scale factors are included.\label{fig:detado}}
\end{figure}

The angular dependence of the missed backscattered fraction can be
determined by integrating over energy the data of Fig.~\ref{fig:120}.
The result is shown in Fig.~\ref{fig:detado}.
A linear fit based on the first 20 keV of data above the analysis cut
was used to extrapolate to 50 eV (the defined threshold for secondary
electrons), so that these integrals and subsequent integrals could be
compared with the current integration measurements.  In order to
estimate the reliability of the fit, an additional systematic
uncertainty was assigned to the extrapolation, based on comparison of
this fit to a fit constrained to pass through zero at zero energy.
For 124 keV beam energy, this extra systematic uncertainty was of
order a few percent.

In order to better estimate the systematic uncertainty in this
extrapolation, simulations were performed using Geant4 and Penelope in
the unmeasured region 50 eV to 20 keV.  In the Geant4 simulation, the
backscattered fraction was found to tend toward zero at small
backscattered electron energy, in fair agreement with the linear
extrapolation method.  In the Penelope simulation, the backscattered
fraction was found to rise steadily as the threshold of the simulation
was reduced.  In the range 100 eV to 10 keV, Penelope gave about 10\%
extra contribution to the integral compared to linear extrapolation,
for beryllium, and an extra 4\% for silicon.  To average between the
extrapolations implied by Geant4 and Penelope, an additional 5\%
contribution was added to the beryllium data, and an additional 2\%
contribution was added to the silicon data.  The additional
uncertainty in each case was taken to be the size of the correction.

\begin{figure}
\includegraphics[width=0.5\textwidth]{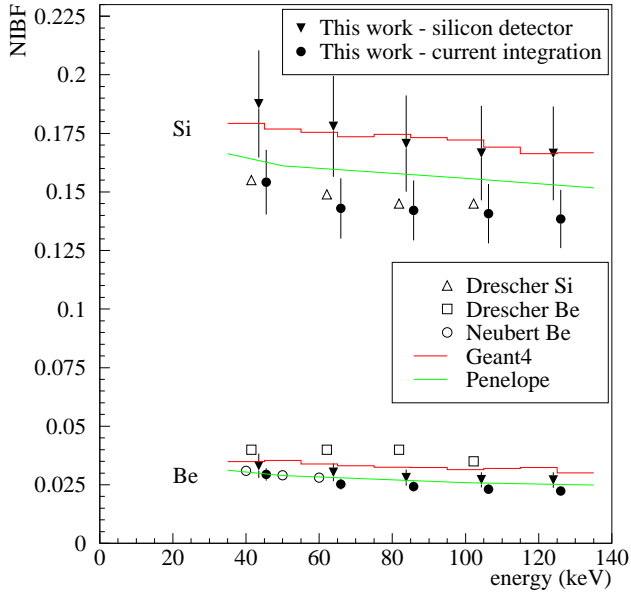}
\caption{(Color online) Normal incidence backscattering from beryllium
and silicon targets.  Integrated silicon detector measurements are
shown by the inverted filled triangles.  Current integration
measurements are shown by filled circles.  Total systematic
uncertainties are shown and the current integration measurements are
displaced by 2 keV so that the error bars do not overlap.  Previous
current integration measurements due to Drescher et
al.~\cite{bib:Drescher} and Neubert et al.~\cite{bib:Neubert1} are
displayed.  The histograms show the results of the Geant4 and Penelope
Monte Carlo simulations with no Monte Carlo scale factor.
\label{fig:nibf}}
\end{figure}

The data were integrated over angle, using the same method described
in relation to Fig.~\ref{fig:be50deg}, to determine the total
normal-incidence backscattered fraction.  The results of this
integration are shown in Fig.~\ref{fig:nibf} and are compared with
current integration measurements (described in
section~\ref{sec:current}) and Monte Carlo simulations (described in
Section~\ref{sec:MC}).
In Fig.~\ref{fig:nibf}, the total systematic uncertainty, including
extrapolation to 50~V and extrapolation over unmeasured angles, is
plotted.

\subsection{Current Integration Mode\label{sec:current}}

The results for $\eta$ based on our current integration measurements
are also shown in Fig.~\ref{fig:nibf}.

The silicon detector measurements are found to be systematically
higher than the current-mode measurements; however, the two methods of
are found to agree within the systematic uncertainties.  In the case
of the current integration method, this systematic uncertainty is
dominated by residual correction for secondary electron collection due
to the penetration through the grid of the target rod.  In the case of
the silicon detector measurements, it is dominated by reproducibility
of the measurements under varying conditions, and by uncertainties in
alignment and solid angle effects.

The data are also compared with previous data on Be and Si targets due
to Drescher et al.~\cite{bib:Drescher} and with data on Be due to
Neubert et al.~\cite{bib:Neubert1}.  Both groups used current
integration techniques to arrive at their results.  Neubert et
al.~\cite{bib:Neubert1} in particular used a second target apparatus
to study the effects of secondary electrons, as opposed to the grid
used in this work.  Only the subsets of their data that overlap the
region 43.5 to 124 keV are plotted.  The Drescher data on Be are
systematically higher than the Neubert data.  However, due to the lack
of additional data on this element, it is impossible to say which is
more accurate.  Our data tend to agree with the Neubert data, as do
the data of Massoumi et al.~\cite{bib:massoumi1,bib:massoumi2} taken
below 40 keV.

Our data on Si is in good agreement with those of Drescher, and tend
to suggest that there could be some systematic effect in either the
current integration or silicon detector data for that element.  Given
the current level of the systematic uncertainties, it is difficult to
make a firm statement.

\section{Comparison with Monte Carlo\label{sec:MC}}

Monte Carlo simulations were conducted using the Geant4 Monte Carlo
\cite{bib:geant4nim} and the Penelope Monte Carlo \cite{bib:penelope2}.

The version of Geant4 used was 4.4.0.  In order to achieve a
reasonable description of backscattering, it was found that three
parameters had to be changed from their default values: the maximum
step size, the threshold to create secondaries, and a parameter
originally introduced into Geant4 to tune low-energy EM processes.
For all our simulations, the maximum step size was set to the range of
a 1~keV electron; the threshold to create secondaries was set to
1~$\mu$m; and the tuning parameter was set to zero.  The step size and
threshold parameters were chosen by reducing them until the integral
backscattered fraction was relatively stable under variation of those
parameters, and to be as small as possible for reasonable running
time.  The threshold parameter additionally was checked to give good
results at low energy for thin targets.  The tuning parameter had to
be changed to zero, as it had been found to be erroneously set to 1.5
in this version of Geant4.  Examples of how to make these
modifications were supplied by the Geant4 Electromagnetic Physics and
Low-Energy Electromagnetic physics groups~\cite{bib:urban}.

The version of Penelope used was 2002b.  Penelope was studied in
detail under the variation of several simulation parameters and was
found to be stable.  The simulation parameters were therefore chosen
to optimize simulation speed consistent with a full detailed
simulation.  For the comparison with our silicon detector data, the
most suitable simulation parameters were found to be: $E_{abs}=10~{\rm
keV}$, $W_{cc}=W_{cr}=5~{\rm keV}$, DSMAX~=~0.005~cm, and
$C_1=C_2=0.05$~\cite{bib:seth}.  These parameters control energy
cut-offs, the maximum step size, and the description of elastic and
inelastic scattering in the medium.  Particles identified as
secondaries by Penelope were included.  Secondaries were also studied
in separate simulations related to our current-integration
measurements, as mentioned earlier.

In both Monte Carlo models, backscattering is defined as any electron
which exits the surface of the target.  The Monte Carlo simulations
included silicon detector response in a simple model including
normal-incidence backscattering from the front face.  This was
validated by the fact that the detector response to mono-energetic
electrons did not vary with beam position on the detector face, as
discussed in Section~\ref{sec:syst}.  Energy smearing based on a 2.5
keV energy-independent energy resolution of the silicon detector was
also included.  Chamber background effects were studied in separate
Penelope-based simulations, as described earlier.

The main difference between Penelope and Geant4 relevant to
backscattering is in the treatment of scattering from nuclei.
Penelope treats these as Rutherford scattering events exactly using a
relativistic screened Rutherford scattering cross section.  Multiple
electron-atom scattering events, with energy loss (dominated by
electron-electron interactions in the target) are found to dominate
backscattering at these energies in this model, for thick targets.
Geant4, on the other hand, has no exact treatment of Rutherford
scattering, relying on sampling from a multiple scattering
distribution.

Fig.~\ref{fig:120} compares the Geant4-based Monte Carlo with our
silicon detector measurements at 124 keV.  It can be seen that Geant4
somewhat overestimates the beryllium data, while having relatively
good agreement with the silicon data.  The beryllium data is globally
overestimated by roughly 20\%, hence a scale factor of 0.8 was applied
to the Monte Carlo for this comparison so that the differences between
the distributions can be more easily seen.  In all cases, it is also
apparent that the peak near $q=0.95$ is systematically underestimated.
However the positions of the low-energy and elastic peaks are rather
well-described by the Monte Carlo.  In the case of a silicon target,
the Geant4 low-energy peak is at a slightly higher $q$ than the data,
while for beryllium this peak appears at slightly lower $q$ relative
to the data.

\begin{figure}
\includegraphics[width=0.5\textwidth]{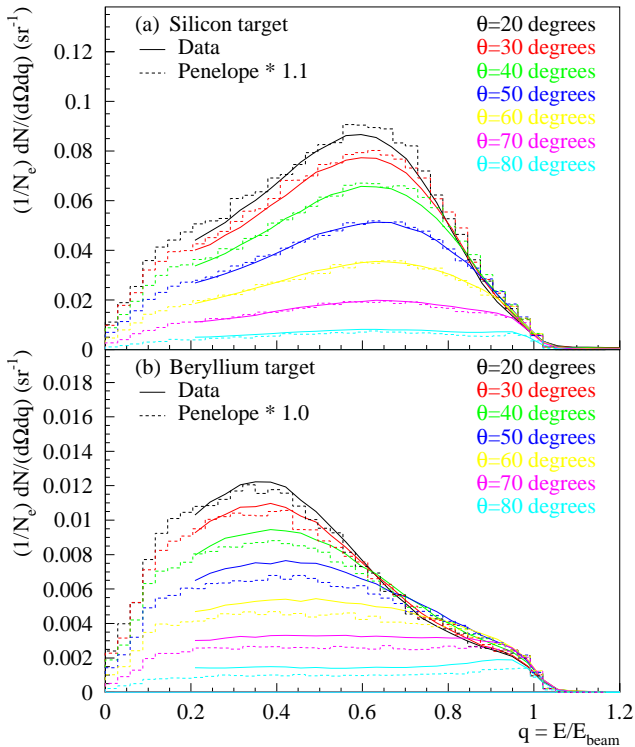}
\caption{(Color online) Normal incidence backscattering from
a. silicon and b. beryllium targets at $E_{\rm beam}=124$~keV.  Curves
represent rebinned data taken with silicon detector.  Histogram is
Monte Carlo simulation based on Penelope.  Systematic uncertainty in
the normalization of the data is estimated to be 12\%.  For silicon, a
scale factor of 1.1 is globally applied to the Penelope simulation.
For beryllium no scale factor is applied.\label{fig:120pen}}
\end{figure}

Fig.~\ref{fig:120pen} compares Penelope-based Monte Carlo with our
measurements using the silicon detector.
A scale factor of 1.1 is applied to the Penelope simulation in the
case of silicon.  No scaling is applied to the beryllium simulation.
The Penelope simulation in general somewhat underestimates the
beryllium measurements by 5\% and underestimates the silicon
measurements by roughly 10\%.  When the Monte Carlo is rescaled, it is
apparent that trends in both energy and angle are well represented by
Penelope.  Since the systematic uncertainty is larger than these
discrepancies, the overall agreement is good.

Fig.~\ref{fig:detado} shows the Geant4 and Penelope simulations
compared with data for the energy-integrated angular distributions
($d\eta/d\Omega$).  No Monte Carlo scale factors are applied for this
comparison.  The same discrepancies in overall magnitude of
backscattering can be seen.  As noted previously, the Penelope
simulation tends to better describe the angular distribution (aside
from the overall scale factor).  The Geant4 distributions are somewhat
narrower compared to the data and Penelope.  Additionally, the Geant4
simulation gives systematically larger backscattering from each
material than does the Penelope simulation.

Fig.~\ref{fig:nibf} compares the integrated $\eta$ results for
different beam energies with Geant4 and Penelope simulations.  The
same discrepancies in normalization are again observed.  Both Penelope
and Geant4 adequately describe the reduction of $\eta$ as the beam
energy increases.

Comparisons of other Monte Carlo models to Geant4, Penelope, and the
existing backscattering data were also carried
out and were reported elsewhere~\cite{bib:seth}.

\section{Conclusion}

A detailed data set for normal incidence backscattering from beryllium
and silicon bulk targets has been acquired for incident electron
energies from 43.5 to 124 keV.  Two methods of determining the total
normal-incidence backscattered fraction were compared, and found to
agree within systematic uncertainties.  The data agree qualitatively
with previous measurements of the doubly-differential distributions of
backscatter done at lower energy.  They also agree quantitatively with
previous measurements using current integration techniques.  The data
agree well with models implemented in the Geant4 and Penelope Monte
Carlo codes.  In terms of overall normalization, Geant4 is found to
give a good description of the data.  Penelope is found to give good
agreement in terms of both overall normalization and relative
distributions of backscattered electrons in angle and energy.

In future measurements, we plan to investigate non-conducting targets
such as scintillator (particularly interesting for a variety of
nuclear physics applications) and active targets (such as scintillator
and silicon detectors).  We also plan to extend this data set to
higher energy (up to 1 MeV) using the same setup with a higher energy
electron source.  This will allow further tests of the available Monte
Carlo models.  The comparison of these data to Monte Carlo models will
provide an important constraint on a part of the systematic
uncertainty for future high precision measurements of neutron beta
decay.  Future work will also include investigation of thin films and
oblique-incidence backscattering which are both important in
understanding systematic effects in these experiments.

We gratefully acknowledge the technical support of Robert Carr.  This work 
was supported by the National Science Foundation.

\bibliography{bs_paper}

\begin{thebibliography}{16}
\expandafter\ifx\csname natexlab\endcsname\relax\def\natexlab#1{#1}\fi
\expandafter\ifx\csname bibnamefont\endcsname\relax
  \def\bibnamefont#1{#1}\fi
\expandafter\ifx\csname bibfnamefont\endcsname\relax
  \def\bibfnamefont#1{#1}\fi
\expandafter\ifx\csname citenamefont\endcsname\relax
  \def\citenamefont#1{#1}\fi
\expandafter\ifx\csname url\endcsname\relax
  \def\url#1{\texttt{#1}}\fi
\expandafter\ifx\csname urlprefix\endcsname\relax\def\urlprefix{URL }\fi
\providecommand{\bibinfo}[2]{#2}
\providecommand{\eprint}[2][]{\url{#2}}

\bibitem[{\citenamefont{Massoumi et~al.}(1992)}]{bib:massoumi1}
\bibinfo{author}{\bibfnamefont{G.~R.} \bibnamefont{Massoumi}}
  \bibnamefont{et~al.}, \bibinfo{journal}{Phys. Rev. Lett.}
  \textbf{\bibinfo{volume}{68}}, \bibinfo{pages}{3873} (\bibinfo{year}{1992}).

\bibitem[{\citenamefont{Massoumi et~al.}(1993)}]{bib:massoumi2}
\bibinfo{author}{\bibfnamefont{G.~R.} \bibnamefont{Massoumi}}
  \bibnamefont{et~al.}, \bibinfo{journal}{Phys. Rev.}
  \textbf{\bibinfo{volume}{B47}}, \bibinfo{pages}{11007}
  (\bibinfo{year}{1993}).

\bibitem[{\citenamefont{G\'erard et~al.}(1995)}]{bib:gerard}
\bibinfo{author}{\bibfnamefont{P.}~\bibnamefont{G\'erard}}
  \bibnamefont{et~al.}, \bibinfo{journal}{Scanning}
  \textbf{\bibinfo{volume}{17}}, \bibinfo{pages}{377} (\bibinfo{year}{1995}).

\bibitem[{\citenamefont{Agostinelli et~al.}(2003)}]{bib:geant4nim}
\bibinfo{author}{\bibfnamefont{S.}~\bibnamefont{Agostinelli}}
  \bibnamefont{et~al.} (\bibinfo{collaboration}{The Geant4 Collaboration}),
  \bibinfo{journal}{Nucl. Instrum. Meth.} \textbf{\bibinfo{volume}{A506}},
  \bibinfo{pages}{250} (\bibinfo{year}{2003}).

\bibitem[{\citenamefont{T.~Tabata and Shinoda}(1999)}]{bib:Tabata2}
\bibinfo{author}{\bibfnamefont{P.~A.} \bibnamefont{T.~Tabata}}
  \bibnamefont{and} \bibinfo{author}{\bibfnamefont{K.}~\bibnamefont{Shinoda}},
  \bibinfo{journal}{Radiation Physics and Chemistry}
  \textbf{\bibinfo{volume}{54}}, \bibinfo{pages}{11} (\bibinfo{year}{1999}).

\bibitem[{\citenamefont{Bowles and {Young co-principal
  investigators}}(2000)}]{bib:ucna}
\bibinfo{author}{\bibfnamefont{T.}~\bibnamefont{Bowles}} \bibnamefont{and}
  \bibinfo{author}{\bibfnamefont{A.~R.} \bibnamefont{{Young co-principal
  investigators}}}, \emph{\bibinfo{title}{A proposal for an accurate
  measurement of the neutron spin, electron angular correlation in polarized
  neutron beta-decay with ultracold neutrons}} (\bibinfo{year}{2000}).

\bibitem[{\citenamefont{Yuan et~al.}(2001)}]{bib:junhua}
\bibinfo{author}{\bibfnamefont{J.}~\bibnamefont{Yuan}} \bibnamefont{et~al.},
  \bibinfo{journal}{Nucl. Instrum. Meth.} \textbf{\bibinfo{volume}{A465}},
  \bibinfo{pages}{404} (\bibinfo{year}{2001}).

\bibitem[{\citenamefont{{The Goodfellow Corporation, 800 Lancaster Avenue,
  Berwyn, PA 19312-1780}}()}]{bib:goodfellow}
\bibinfo{author}{\bibnamefont{{The Goodfellow Corporation, 800 Lancaster
  Avenue, Berwyn, PA 19312-1780}}}.

\bibitem[{\citenamefont{Joy}(1995)}]{bib:joy}
\bibinfo{author}{\bibfnamefont{D.~C.} \bibnamefont{Joy}},
  \bibinfo{journal}{Scanning} \textbf{\bibinfo{volume}{17}},
  \bibinfo{pages}{270} (\bibinfo{year}{1995}).

\bibitem[{\citenamefont{Bronshtein and
  Denisov}(1965{\natexlab{a}})}]{bib:bronshtein2}
\bibinfo{author}{\bibfnamefont{I.}~\bibnamefont{Bronshtein}} \bibnamefont{and}
  \bibinfo{author}{\bibfnamefont{S.~S.} \bibnamefont{Denisov}},
  \bibinfo{journal}{Soviet Physics - Solid State} \textbf{\bibinfo{volume}{7}},
  \bibinfo{pages}{1484} (\bibinfo{year}{1965}{\natexlab{a}}).

\bibitem[{\citenamefont{Bronshtein and
  Denisov}(1965{\natexlab{b}})}]{bib:bronshtein}
\bibinfo{author}{\bibfnamefont{I.~M.} \bibnamefont{Bronshtein}}
  \bibnamefont{and} \bibinfo{author}{\bibfnamefont{S.~S.}
  \bibnamefont{Denisov}}, \bibinfo{journal}{Soviet Physics - Solid State}
  \textbf{\bibinfo{volume}{6}}, \bibinfo{pages}{2106}
  (\bibinfo{year}{1965}{\natexlab{b}}).

\bibitem[{\citenamefont{H.~Drescher and Seidel}(1970)}]{bib:Drescher}
\bibinfo{author}{\bibfnamefont{L.~R.} \bibnamefont{H.~Drescher}}
  \bibnamefont{and} \bibinfo{author}{\bibfnamefont{H.}~\bibnamefont{Seidel}},
  \bibinfo{journal}{Zeit. Angewandte Physik} \textbf{\bibinfo{volume}{29}},
  \bibinfo{pages}{331} (\bibinfo{year}{1970}).

\bibitem[{\citenamefont{Neubert and Rogashewski}(1980)}]{bib:Neubert1}
\bibinfo{author}{\bibfnamefont{G.}~\bibnamefont{Neubert}} \bibnamefont{and}
  \bibinfo{author}{\bibfnamefont{S.}~\bibnamefont{Rogashewski}},
  \bibinfo{journal}{Phys Stat. Sol. (a)} \textbf{\bibinfo{volume}{59}},
  \bibinfo{pages}{35} (\bibinfo{year}{1980}).

\bibitem[{\citenamefont{Sempau et~al.}(1997)}]{bib:penelope2}
\bibinfo{author}{\bibfnamefont{J.}~\bibnamefont{Sempau}} \bibnamefont{et~al.},
  \bibinfo{journal}{Nucl. Instrum. Meth. B} \textbf{\bibinfo{volume}{132}},
  \bibinfo{pages}{377} (\bibinfo{year}{1997}).

\bibitem[{\citenamefont{Urban and Ivantchenko}(2003)}]{bib:urban}
\bibinfo{author}{\bibfnamefont{L.}~\bibnamefont{Urban}} \bibnamefont{and}
  \bibinfo{author}{\bibfnamefont{V.}~\bibnamefont{Ivantchenko}},
  \bibinfo{journal}{private communications}  (\bibinfo{year}{2003}).

\bibitem[{\citenamefont{Hoedl}(2003)}]{bib:seth}
\bibinfo{author}{\bibfnamefont{S.}~\bibnamefont{Hoedl}},
  \bibinfo{journal}{Princeton Ph.D. Thesis (unpublished)}
  (\bibinfo{year}{2003}).

\end{thebibliography}

\end{document}